\documentclass[5p,authoryear]{elsarticle} 
\usepackage[english]{babel} 
\usepackage{lineno}
\usepackage{amsmath}
\usepackage{amssymb}
\usepackage{verbatim}
\usepackage{multirow}
\usepackage{color,graphicx}
\usepackage{bm}
\usepackage{subfigure}
\usepackage{verbatim}
\usepackage{setspace}
\usepackage[bookmarks=true]{hyperref}
\usepackage{ulem}
\usepackage{array}

\providecommand{\Zxhreftb}[1]{Table~\ref{#1}}
\providecommand{\zxhreftb}[1]{Table~\ref{#1}}
\providecommand{\zxhreffig}[1]{Fig.~\ref{#1}}

\providecommand{\leicomment}[1]{\textcolor[rgb]{0.86,0.48,0.17}{\textbf{lei: #1}}}

\modulolinenumbers[1]
\journal{Medical Image Analysis}
\begin{document}
\begin{frontmatter}
\title{Cardiac Segmentation on Late Gadolinium Enhancement MRI: \\A Benchmark Study from Multi-Sequence Cardiac MR Segmentation Challenge}

\author{Xiahai~Zhuang$^{1}$*} \ead[url]{zxh@fudan.edu.cn}
\author{Jiahang~Xu$^{1}$*} \ead[url]{jhxu18@fudan.edu.cn}
\author{Xinzhe~Luo$^{1}$} 
\author{Chen~Chen$^{3}$} 
\author{Cheng~Ouyang$^{3}$} 
\author{Daniel~Rueckert$^{3}$} 
\author{Victor~M. Campello$^{4}$} 
\author{Karim~Lekadir$^{4}$} 
\author{Sulaiman~Vesal$^{5}$} 
\author{Nishant~RaviKumar$^{5}$} 
\author{Yashu~Liu$^{6}$} 
\author{Gongning~Luo$^{6}$}
\author{Jingkun~Chen$^{7}$} 
\author{Hongwei~Li$^{8}$} 
\author{Buntheng~Ly$^{9}$} 
\author{Maxime~Sermesant$^{9}$}
\author{Holger~Roth$^{10}$} 
\author{Wentao~Zhu$^{10}$} 
\author{Jiexiang~Wang$^{11}$} 
\author{Xinghao~Ding$^{11}$} 
\author{Xinyue~Wang$^{12}$} 
\author{Sen~Yang$^{12,13}$} 
\author{Lei~Li$^{1,2}$*} \ead[url]{lilei.sky@sjtu.edu.cn}

\address{$^1$School of Data Science, Fudan University, Shanghai, China\\[0.5ex]
$^2$School of Biomedical Engineering, Shanghai Jiao Tong University, Shanghai, China \\[0.5ex]
$^3$Biomedical Image Analysis Group, Imperial College London, London, UK\\[0.5ex]
$^4$Department Mathematics \& Computer Science, Universitat de Barcelona, Barcelona, Spain\\[0.5ex]
$^5$Friedrich-Alexander-Universit{\"a}t Erlangen-N{\"u}rnberg, Germany\\[0.5ex]
$^6$School of Computer Science and Technology, Harbin Institute of Technology, Harbin, China\\[0.5ex]
$^{7}$Department of Computer Science and Engineering, Southern University of Science and Technology, Shenzhen, China\\[0.5ex]
$^{8}$Department of Informatics, Technical University of Munich, Germany\\[0.5ex]
$^{9}$INRIA, Universit\'{e} C\^{o}te d'Azur, Sophia Antipolis, France\\[0.5ex]
$^{10}$NVIDIA, Bethesda, USA\\[0.5ex]
$^{11}$School of Informatics, Xiamen University, Xiamen, China\\[0.5ex]
$^{12}$College of Electrical Engineering, Sichuan University, Chengdu, China\\[0.5ex]
$^{13}$Tencent AI Lab, Shenzhen, China
}

\begin{abstract}

Accurate computing, analysis and modeling of the ventricles and myocardium from medical images are important, especially in the diagnosis and treatment management for patients suffering from myocardial infarction (MI).
Late gadolinium enhancement (LGE) cardiac magnetic resonance (CMR) provides an important protocol to visualize MI.
However, automated segmentation of LGE CMR is still challenging, due to the indistinguishable boundaries, heterogeneous intensity distribution and complex enhancement patterns of pathological myocardium from LGE CMR. Furthermore, compared with the other sequences LGE CMR images with gold standard labels are particularly limited, which represents another obstacle for developing novel algorithms for automatic segmentation of LGE CMR.
This paper presents the selective results from the Multi-Sequence Cardiac MR (MS-CMR) Segmentation challenge, in conjunction with MICCAI 2019.
The challenge offered a data set of paired MS-CMR images, including auxiliary CMR sequences as well as LGE CMR, from 45 patients who underwent cardiomyopathy.
It was aimed to develop new algorithms, as well as benchmark existing ones for LGE CMR segmentation and compare them objectively.
In addition, the paired MS-CMR images could enable algorithms to combine the complementary information from the other sequences for the segmentation of LGE CMR.
Nine representative works were selected for evaluation and comparisons, among which three methods are unsupervised methods and the other six are supervised.
The results showed that the average performance of the nine methods was comparable to the inter-observer variations. Particularly, the top-ranking algorithms from both the supervised and unsupervised methods could generate reliable and robust segmentation results. The success of these methods was mainly attributed to the inclusion of the auxiliary sequences from the MS-CMR images, which provide important label information for the training of deep neural networks.
The challenge continues as an ongoing resource, and the gold standard segmentation as well as the MS-CMR images of both the training and test data are available upon registration via its homepage (www.sdspeople.fudan.edu.cn/zhuangxiahai/0/mscmrseg/).

\end{abstract}

\begin{keyword}
Multi-Sequence \sep Cardiac MRI Segmentation \sep Benchmark \sep Challenge
\end{keyword}

\end{frontmatter}

\begin{spacing}{1}

\section{Introduction}

\begin{figure*}[t]\center
 \includegraphics[width=1.0\textwidth]{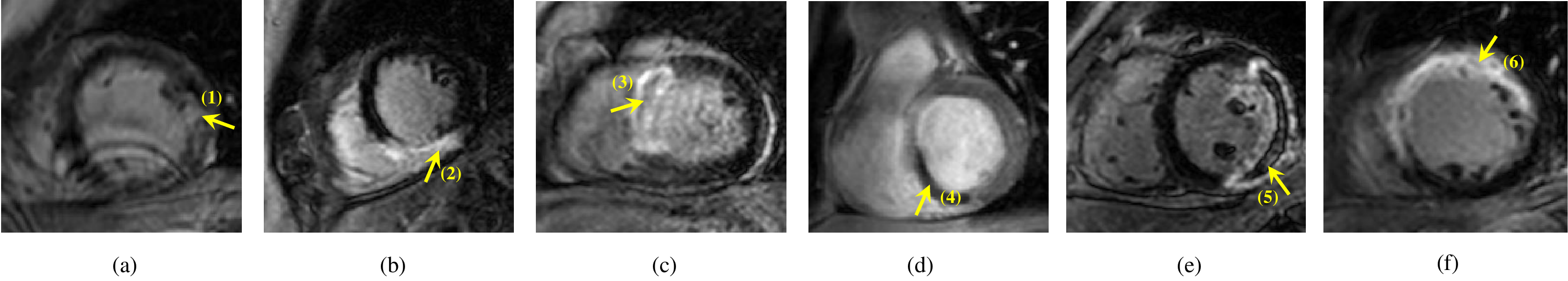}\\[-2ex]
   \caption{Examples of LGE CMR slices, where scars are pointed out by arrows. One can see the enhancement patterns can be complex, and the scars are diverse in location, shape and size. Note that scars in (3), (4) and (5) include MVO appearing dark inside, and the boundary between blood pools and infarcts can be generally indistinct, particularly in (1) and (3).}
\label{fig:intro:challenges}\end{figure*}

Myocardial infarction (MI) is a major cause of deaths and disability worldwide \citep{ journal/EHJ/Thygesen2008}.
Assessment of myocardial viability is essential in the diagnosis and treatment management for patients suffering from MI.
Late gadolinium enhancement (LGE) cardiac magnetic resonance (CMR) sequence can visualize MI,
attributed to the slow washout kinetics of the gadolinium in the infarcted areas,
which appear with distinctive brightness compared with the health tissues.
The sensitivity and accuracy of LGE CMR for the assessment of the presence, location, and extent of MI have been well validated \citep{journal/jacc/amado2004, journal/Lancet/wagner2003}.

Before the delineation of MI, accurate segmentation of myocardium is generally required \citep{journal/JACC/flett2011}.
However, automating this segmentation remains challenging, due to the large shape variability of heart, indistinguishable boundaries, and the possible poor image quality, as \zxhreffig{fig:intro:challenges} shows.
Particularly, there are three challenges for the automatic analysis of LGE CMR.
Firstly, the intensity distribution of the pathological myocardium is heterogeneous.
Hence, it is difficult to model the intensity distribution using parametric models, which usually work well for other CMR sequences.
Secondly, the enhancement patterns can be complex.
The location, shape and size of infarcts vary greatly across different patients, as \zxhreffig{fig:intro:challenges} shows.
Particularly, microvascular obstruction (MVO), which could occur in a high proportion of MI patients, appears particularly dark inside the bright infarct region, due to the lack of uptake of contrast agent.
Thirdly, the intensity distributions of the myocardium and its surroundings are conjoining, making it even more difficult to differentiate the boundaries, such as the boundaries between scars and blood pools and the boundaries between healthy myocardium and liver.

\begin{table*}\center
\caption{Summary of the previous challenges related to cardiac segmentation from MICCAI/ ISBI society.
LV: left ventricle; Myo: myocardium; RV: right ventricle;
LA: left atrium; WHBP: whole heart blood pool; WH: whole heart.}
\label{tb:table:challenge}
{\small
\begin{tabular}{ lllll} \hline
Reference	   &  Year & Data & Target & Pathologies \\
\hline
\citet{journal/ij/radau2009} 					& 2009 & 45 bSSFP MRI		  & LV, Myo				& hypertrophy, infarction\\
\citet{conf/stacom/Suinesiaputra2011} & 2011 & 200 bSSFP MRI		 & LV, Myo					& myocardial infarction \\
\citet{journal/MedAI/Petitjean2015}   	& 2012 & 48 bSSFP MRI		  & RV					& congenital heart disease \\
\citet{journal/MedAI/karim2016}	   	& 2012 & 30 LGE MRI		& LV scars			   & myocardial infarction\\
\citet{journal/jcmr/Karim2013}		 		& 2013 & 60 LGE MRI		   & LA scar			   & atrial fibrillation\\
\citet{journal/tmi/Tobon2015}		 		& 2013 & 30 CT, 30 bSSFP MRI		& LA					& atrial fibrillation\\
\citet{journal/MedAI/karim2018}	   	& 2016 & 10 CT, 10 black-blood MRI		& LA wall			   & atrial fibrillation\\
\citet{link/HVSMR2016}						& 2016 & 20 bSSFP MRI		  & WHBP, Myo	   & congenital heart disease \\
\citet{journal/tmi/bernard2018}	   	& 2017 & 150 bSSFP MRI		 & LV, Myo, RV		   & infarction, dilated/ hypertrophic \\[-0.2ex]
  & & & &cardiomyopathy, abnormal RV\\
\citet{journal/MedAI/zhuang2019}	& 2017 & 60 CT, 60 bSSFP MRI & WH				   & atrial fibrillation, congenital heart\\[-0.2ex]
  & & & & disease, coronary heart disease\\
\citet{link/LAseg2018}						& 2018 & 150 LGE MRI		   & LA					& atrial fibrillation \\

\hline
\end{tabular}}\\
\end{table*}

To the best of our knowledge, limited work has been done for the fully automatic segmentation from LGE CMR,
in contrast to the rich literature of cardiac segmentation on the balanced-Steady State Free Precession (bSSFP) CMR.

\subsection{State-of-the-art of LGE CMR segmentation}

Fully automated segmentation of LGE CMR could either solely rely on the information of the LGE CMR themselves, or combine the complementary information from other sequences, such as bSSFP and T2 CMR.
For the former scheme, shape models or constraints are generally required.
\citet{journal/mia/Wei2013} proposed a 1D parametric model to detect the paired endocardial (Endo) and epicardial (Epi) edge points. To maintain a realistic shape, they imposed a thickness constraint on the segmented myocardium for the 3D deformation.
\citet{journal/TBME/liu2017} proposed the multi-component Gaussian mixture model to deal with the intensity heterogeneity of myocardium in the LGE CMR images.
To maintain a realistic shape of the resulting segmentation, they further employed a coupled level set to model the paired Endo and Epi surfaces, and regularize the shape of segmented myocardium to resemble a flat donut on 2D slices.
Recently, \citet{conf/MICCAI/yue2019} proposed a deep neural network (DNN) based method to segment LGE CMR, where the shape prior and spatial prior networks, respectively referred to as shape reconstruction neural network and spatial constraint network, were employed to improve the segmentation results.

\begin{figure}[t]\center
	\includegraphics[width=0.48\textwidth]{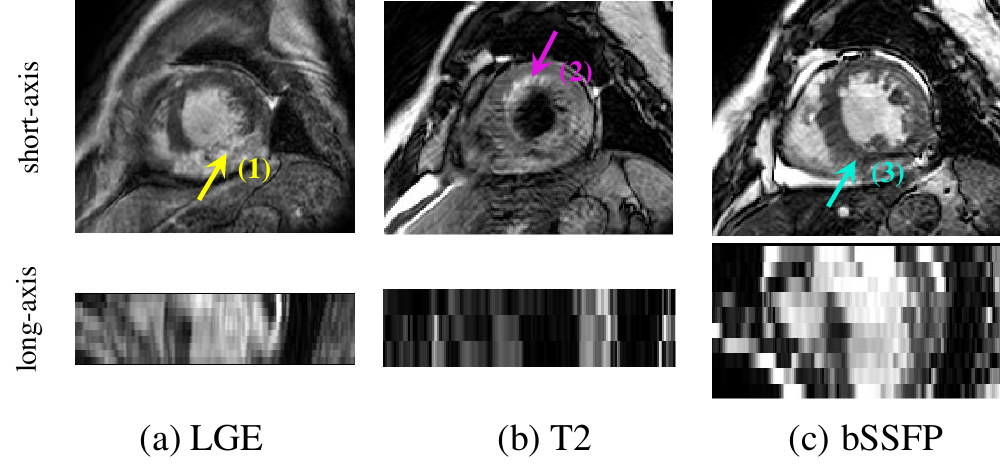}\\[-2ex]
	\caption{Illustration of multi-sequence CMR images acquired from the same patient: (a) LGE CMR visualizes the scars as brighter texture, in contrast to the dark healthy myocardium, as arrow (1) donates; (b) T2 CMR presents the myocardial edema in higher intensity values, as arrow (2) shows, and can visualize the trabeculations and papillary muscles in great detail; (c) bSSFP CMR provides good boundary information between the myocardium and blood pools which are generally indistinct in LGE CMR containing scars, as arrows (1) and (3) indicate.}
	\label{fig:intro:MS-CMR}\end{figure}

Integrating the information from MS-CMR sequences, particularly from the same patient, can assist the segmentation of LGE CMR \citep{journal/pami/Zhuang2019}.
This is because the appearance of the myocardium could be clearer in other CMR sequences.
For example, both T2-weighted CMR and bSSFP cine sequence present clearer boundaries of myocardium.
\zxhreffig{fig:intro:MS-CMR} gives an example of the three sequences from our challenge.
A straightforward solution is to propagate the segmentation of the auxiliary sequence(s), such as bSSFP CMR acquired from the same subject and the same session, directly to the target LGE CMR \citep{conf/miccai/Dikici04,journal/jcmr/Lu2013,journal/jmri/Tao2015}.
This requires accurate segmentation from bSSFP CMR, generally done by manual work, and a robust registration, to avoid accumulated errors.

Recently, more advanced strategies have been adopted to maintain an accurate and shape-realistic segmentation on LGE CMR.
The strategies include (1) transferring the shape prior knowledge from bSSFP CMR to the segmentation of LEG CMR and (2) simultaneous segmentation of MS-CMR images.
For transferring shape prior, 
\citet{journal/CARS/liu2018} built a sparse shape model and target-specific dictionary from a set of bSSFP CMR images. The cardiac segmentation of an unseen LGE CMR image was then sparsely recomposed from the shape dictionary.
For the simultaneous segmentation combining MS-CMR, \citet{conf/ISBI/liu2014} and \citet{journal/cmig/liu2019} combine the information of LGE and T2 images from the same subject to obtain a shape-constrained segmentation result of the LGE CMR.
\citet{conf/miccai/Zhuang16,journal/pami/Zhuang2019} proposed a multivariate mixture model (MvMM) to combine the information of three-sequence CMR from the same subject. The three-source images were then simultaneously registered and segmented within a unified framework.
These works have illustrated the great potential of combining multi-source images for the automatic segmentation of LGE CMR.


\subsection{Motivation}\label{motivation}
For LGE CMR, not only the papers focusing on automated segmentation are scarce,
but also the open data are rarely released.
\zxhreftb{tb:table:challenge} presents the recent challenges and public datasets for cardiac segmentation.
One can see that only \citet{journal/MedAI/karim2016}, \citet{journal/jcmr/Karim2013} and \citet{link/LAseg2018} provided the LGE CMR images for benchmark studies, of which two were for left atrial segmentation and one for ventricles. None of them includes multi-source images to assist the segmentation of LGE CMR.

We therefore organized the MS-CMR Segmentation challenge 2019, consisting of 45 MS-CMR images, in conjunction with MICCAI 2019.
The challenge particularly provided three-sequence CMR, including the LGE CMR, and was aimed to encourage the development of new segmentation algorithms, as well as validating existing ones.
Twenty-three submissions were evaluated before the deadline, and fourteen teams presented their work at the conference event.
In this paper, we introduce the related information of the challenge, elaborate on the methodologies of the representative methods, and analyze their results in details.

The rest of this paper is organized as follows.
Section \ref{material} provides details of the materials and evaluation framework from the challenge.
Section \ref{methods} introduces the evaluated methods for benchmarking.
Section \ref{result} presents the results, followed by discussions in Section \ref{discussion}.
Finally, we conclude this work in Section \ref{conclusion}.

\section{Materials and setup} \label{material}

\subsection{MS-CMR images}\label{material:data}

MS-CMR segmentation challenge provided image sets from 45 subjects who underwent cardiomyopathy, of which each consists of three CMR sequences, i.e. LGE CMR, bSSFP CMR and T2 CMR \citep{conf/miccai/Zhuang16,journal/pami/Zhuang2019}.
The data had been collected with institutional ethics approval and had been anonymized.

The three CMR sequences were all breath-hold, multi-slice, acquired from the cardiac short-axis views.
The LGE CMR was a T1-weighted, inversion-recovery, gradient-echo sequence, consisting of 10 to 18 slices and covering the main body of the ventricles.
The typical parameters are as follows, TR/TE: 3.6/1.8 ms; slice thickness: 5 mm; in-plane resolution: reconstructed into 0.75 $\!\times\!$ 0.75 mm.

The T2 CMR was a T2-weighted, black blood Spectral Presaturation Attenuated Inversion-Recovery (SPAIR) sequence, generally consisting of a small number of slices.
For example, among the 35 cases, 13 have only three slices, and the others have five (13 subjects), six (8 subjects) or seven (one subject) slices.
The typical parameters are as follows, TR/TE: 2000/90 ms; slice thickness: 12-20 mm; in-plane resolution: reconstructed into 1.35 $\!\times\!$ 1.35 mm.

The bSSFP CMR was a balanced-steady state free precession cine sequence.
Since both the LGE and T2 CMR were scanned at the end-diastolic phase, the same cardiac phase of the bSSFP cine data was selected for this study.
The bSSFP images generally consist of 8 to 12 contiguous slices, covering the full ventricles from the apex to the basal plane of the mitral valve, with some cases having several slices beyond the ventricles.
The typical parameters are as follows, TR/TE: 2.7/1.4 ms; slice thickness: 8-13 mm; inplane resolution: reconstructed into 1.25 $\!\times\!$ 1.25 mm.

The data were split into two sets, i.e., the training set and the test set, as \zxhreftb{tb:material:dataset} shows.
For the training data, both the LGE CMR images and the corresponding gold standard were released to the participants for building, training and cross-validating their models.
Besides, the challenge provides images and gold standard labels from the other two sequences, to assist the segmentation of LGE CMR.
For the test data, only LGE CMR images were released.
Once the participants developed their algorithms, they could submit their segmentation results of the test data to the challenge moderators for an independent evaluation.
To avoid parameter tuning, the challenge only allowed a maximum of two evaluations for one algorithm.
\emph{Note that the MS-CMR data and gold standard segmentation of all images are now available upon request and registration via the challenge homepage\footnote{www.sdspeople.fudan.edu.cn/zhuangxiahai/0/mscmrseg19/}.}

\subsection{Gold standard segmentation and evaluation metrics}

Three observers were employed to manually label the left ventricular blood pool (LV), right ventricular blood pool (RV) and left ventricular myocardium (Myo).
The observers were well trained raters who were post-graduate students either in biomedical engineering or medical imaging field.
The manual labeling was performed slice-by-slice using a brush tool in the software ITK-SNAP \citep{journal/Neuroimage/Yushkevich2006}.
On average, each of the segmentation took 20 minutes.
All the manual segmentation results were validated by senior experts in cardiac anatomy before used in the construction of gold standard segmentation,
which was achieved by averaging the three manual delineations using the shape-based approach \citep{journal/tip/rohlfing2006}.

To evaluate the accuracy of a segmentation result, we employed two widely used metrics, i.e.,
the Dice score and Hausdorff Distance (HD) \citep{journal/jhe/Zhuang13},
which are respectively defined as,
\begin{equation}
  \mathrm{Dice}(V_{\mathrm{seg}},V_{\mathrm{GD}}) = \frac{2\left|V_{\mathrm{seg}} \cap V_{\mathrm{GD}}\right|}{\left|V_{\mathrm{seq}}\right|+\left|V_{\mathrm{GD}}\right|},
\end{equation}
and
\begin{equation}
  \mathrm{HD}(X, Y)=\max \Big[\sup _{x \in X} \inf _{y \in Y} d(x, y), \sup _{y \in Y} \inf _{x \in X} d(x, y)\Big],
\end{equation}
where $V_{\mathrm{GD}}$ and $V_{\mathrm{seq}}$ denote the gold standard and automatic segmentation, respectively;
$X$ and $Y$ represent two sets of contour points, and $d(x, y)$ indicates the distance between the two points $x$ and $y$.

\begin{table} [t] \center
\caption{The distribution of training and test data. $I^\mathrm{lab}$: images with gold standard labels; $I^\mathrm{unl}$: images without labels.
} \label{tb:material:dataset}
{\small
\begin{tabular}{l| lllll}\hline
Dataset & LGE & T2 &   bSSFP\\
\hline
Training &  5 $I^\mathrm{lab}$ &  35 $I^\mathrm{lab}$, 10 $I^\mathrm{unl}$ &  35 $I^\mathrm{lab}$, 10 $I^\mathrm{unl}$ \\
Test  &  40 $I^\mathrm{unl}$ & None & None \\  
\hline
\end{tabular} }\\
\end{table}

\subsection{Participants}
As an ongoing event, the challenge had received eighty-seven requests of registration before the submission of this manuscript, among which sixty-five teams participated the event before the date of the workshop (Oct 13th, 2019).
Twenty-three submitted results were evaluated before the submission deadline, and nine algorithms were selected for this benchmark work.
All these selected teams agreed to include their methods and results for publication in this paper.
The selection criterion was based on the novelty of the proposed methodologies and performance of the algorithms.
\emph{Note that the team abbreviations in the remaining of this paper refer both to the teams and their corresponding methods.}

\begin{table*} [t] \center
	\caption{
	Summary of the benchmark methods. The abbreviations are as follows, Dim: the segmentation is performed on 2D slices or 3D images; DAug: whether use data augmentation strategy; Img Syn: whether use image synthesis or reconstruction strategy; Pre/ Post: whether use pre-processing or post-processing strategy. For pre-processing, R: image resample or resize; C: image cropping; IN: image intensity normalization.
	}
\label{tb:method:summary1}
{\footnotesize
\begin{tabular}{ l| l l l l l | l l l *{7}{@{\ \,}  l}}\hline
\multirow{2}*{Teams}   & \multirow{2}*{Dim}   & \multirow{2}*{DAug}  &  \multirow{2}*{Img Syn} & \multirow{2}*{Network} & \multirow{2}*{Pre/ Post} & \multicolumn{3}{c}{Training data}\\
\cline{7-9}
~   & ~  & ~ & ~ & ~ & ~ & LGE & bSSFP & T2\\
\hline
ICL		& 2D & Y & Y & UNIT + cascaded U-Net	& R, C, IN/ Y & 40 $I^\mathrm{unl}$ & 20 $I^\mathrm{lab}$, 20 $I^\mathrm{unl}$& 0\\
XMU	 & 2D & N & N & Attention U-Net with GFRM & R, C/ N & 40 $I^\mathrm{unl}$ & 35 $I^\mathrm{lab}$ & 35 $I^\mathrm{lab}$\\
INRIA   & 3D & Y & Y & TCL-Net				   & R, IN/ N & 5 $I^\mathrm{unl}$ & 35 $I^\mathrm{lab}$   & 35 $I^\mathrm{lab}$\\
\hline
SCU	 & 2D & Y & N & SK-Unet				   & C, IN/ Y & 5 $I^\mathrm{lab}$	& 35 $I^\mathrm{lab}$ & 35 $I^\mathrm{lab}$ \\
UB	 & 2D & Y & Y & CycleGAN + U-Net		  & R, C, IN/ N & 5 $I^\mathrm{lab}$, 40 $I^\mathrm{unl}$ & 35 $I^\mathrm{lab}$, 10  $I^\mathrm{unl}$ & 35 $I^\mathrm{lab}$\\
FAU	 & 2D & Y & N & U-Net					 & IN, C/ Y & 4 $I^\mathrm{lab}$	& 35 $I^\mathrm{lab}$ & 35 $I^\mathrm{lab}$\\
NVIDIA  & 3D & Y & N & MAS + AH-Net			  & R/ Y & 5 $I^\mathrm{lab}$, 40 $I^\mathrm{unl}$ & 35 $I^\mathrm{lab}$ & 35 $I^\mathrm{lab}$\\
HIT	 & 2D & Y & Y & Res-UNet				  & R, C/ N & 4 $I^\mathrm{lab}$, 30 $I^\mathrm{unl}$ & 35 $I^\mathrm{lab}$ &0 \\
SUSTech	& 2D & Y & N & U-Net with discriminator	 & C, IN/ N & 3 $I^\mathrm{lab}$	 & 35 $I^\mathrm{lab}$ & 35 $I^\mathrm{lab}$\\
\hline
\end{tabular} }
\end{table*}

\section{Evaluated Methods} \label{methods}
In this section, we elaborate on the nine benchmarked algorithms.
Three of them, i.e., ICL, XMU and INRIA, are based on unsupervised learning, without using labeled LGE images.
The other six teams, i.e.,
SCU, UB, FAU, NVIDIA, HIT and SUSTech, trained their models using 3 to 5 labeled LGE images, and thus are considered as weakly supervised methods.
\Zxhreftb{tb:method:summary1} summarizes these algorithms and their training data.

\begin{figure*}[t]\center
	\subfigure[] {\includegraphics[width=0.46\textwidth]{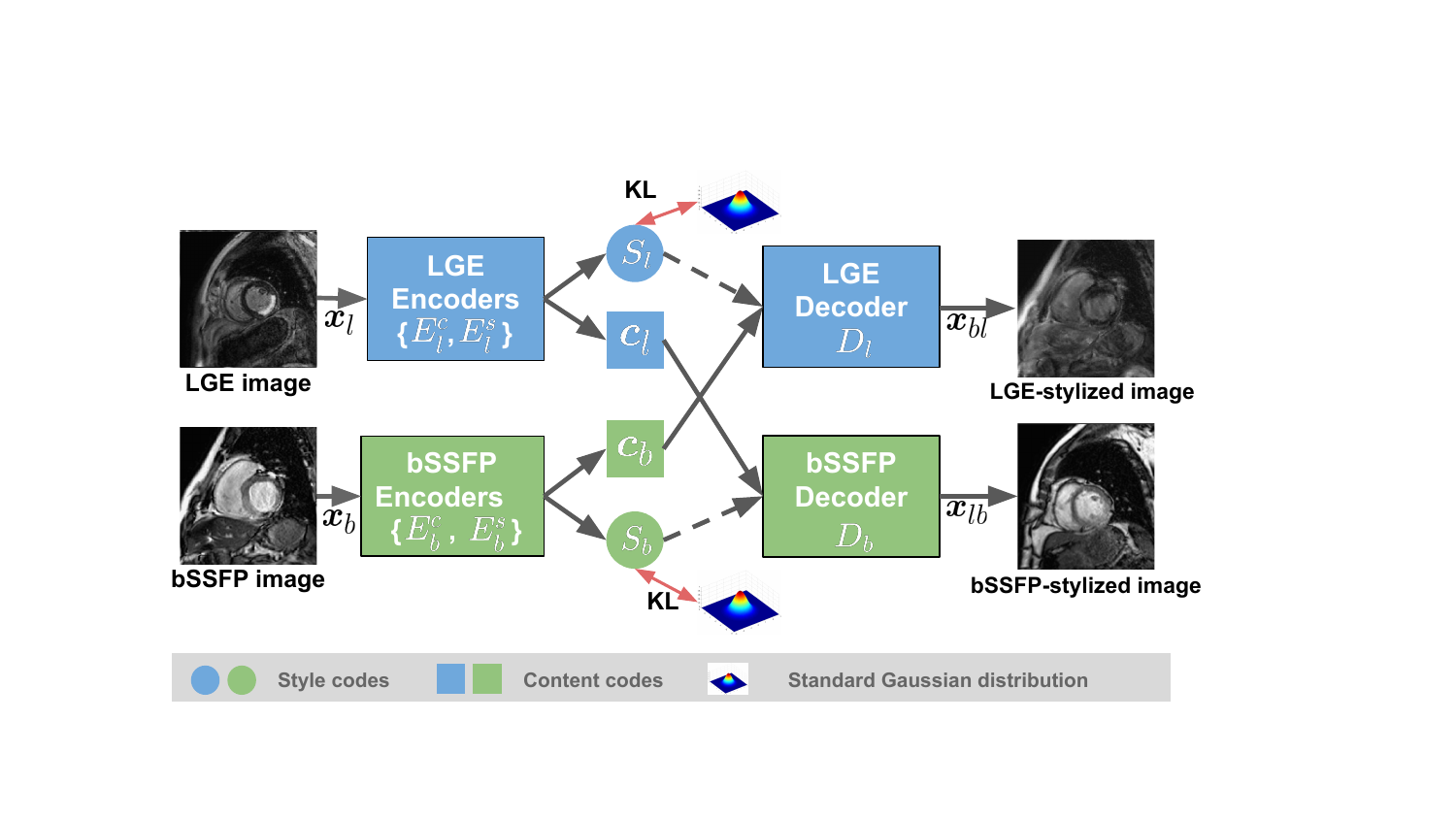}}
	\subfigure[] {\includegraphics[width=0.51\textwidth]{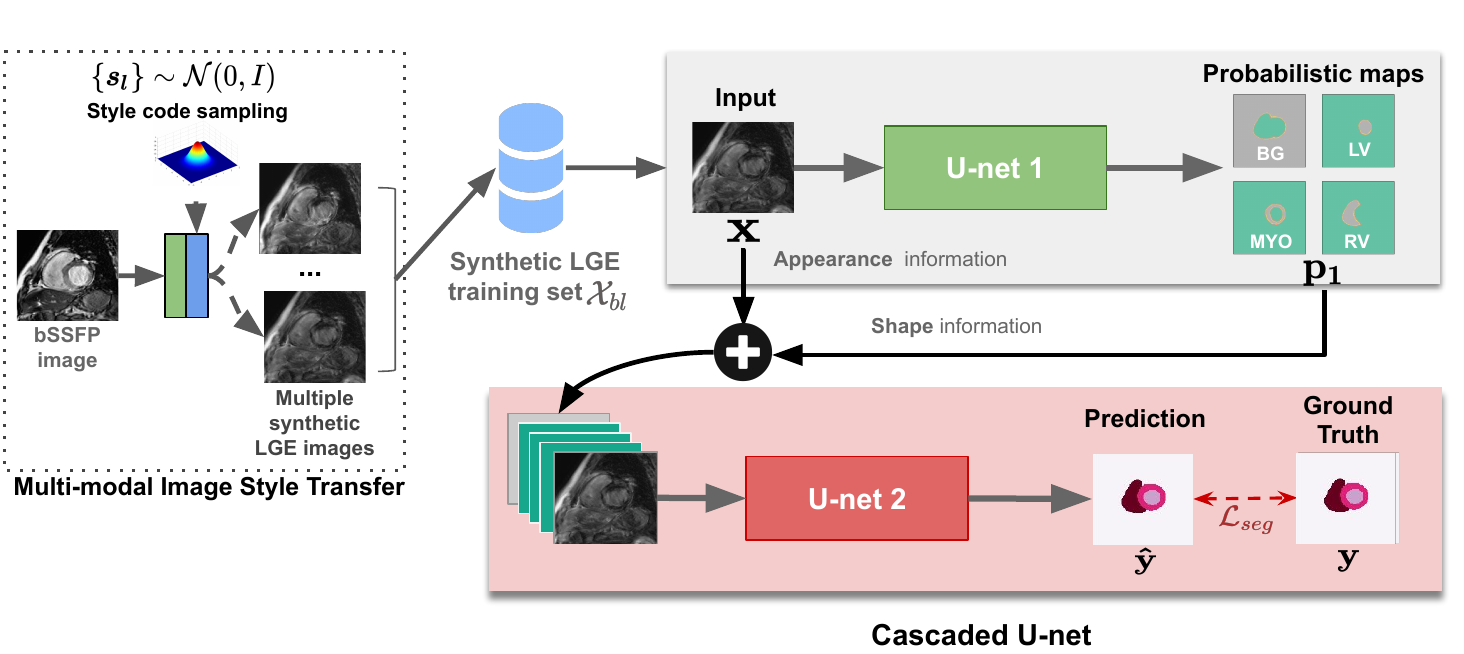}}\\[-2.0ex]
	\caption{Overview of the two-stage framework proposed by ICL:
		(a) the multi-modal image UNIT translation network, which consists of two encoder-decoder pairs for the two domains: bSSFP and LGE, respectively;
		(b) the two-stage cascaded segmentation network.}
	\label{fig:method:ICL}\end{figure*}

\subsection{Imperial College London (ICL)} \label{method:ICL}

ICL formulates an unsupervised learning algorithm, by decomposing the training process into two stages, i.e. image translation for style transfer from bSSFP to LGE and image segmentation \citep{conf/stacom/chen2019_IC}.

As \zxhreffig{fig:method:ICL} shows, the translation network builds on previous work of multi-modal unsupervised image-to-image translation (UNIT) network \citep{conf/ECCV/huang2018}.
Here, the representations from shape and appearance of the two modalities are disentangled, swapped and combined to synthesize LGE-stylized images, which are then fed as training samples to the downstream segmentation network.
Apart from weighted cross-entropy loss, $L_2$ distance loss between edge maps is also employed for contour constraint of cardiac structures, especially the myocardium.


\subsection{Xiamen University (XMU)} \label{method:XMU}
XMU focuses on the alignment of features that are extracted from source (bSSFP and T2) and target (LGE) by two parameter-sharing segmentation networks \citep{conf/STACOM/wang2019}.
The networks are built upon a baseline model that exploits techniques of pyramid pooling and attention modules to achieve learning of fine-grained features.
At the same time, two discriminator networks, minimizing discrepancy in both the feature and output space, are designated for domain adaption from source to target via adversarial training.
The networks are optimized based on a hybrid loss function that incorporates cross entropy and Jaccard loss for segmentation in the source domain, and an adversarial loss for feature and mask alignment in the target domain.

\begin{figure}[t]\center
 \includegraphics[width=0.5\textwidth]{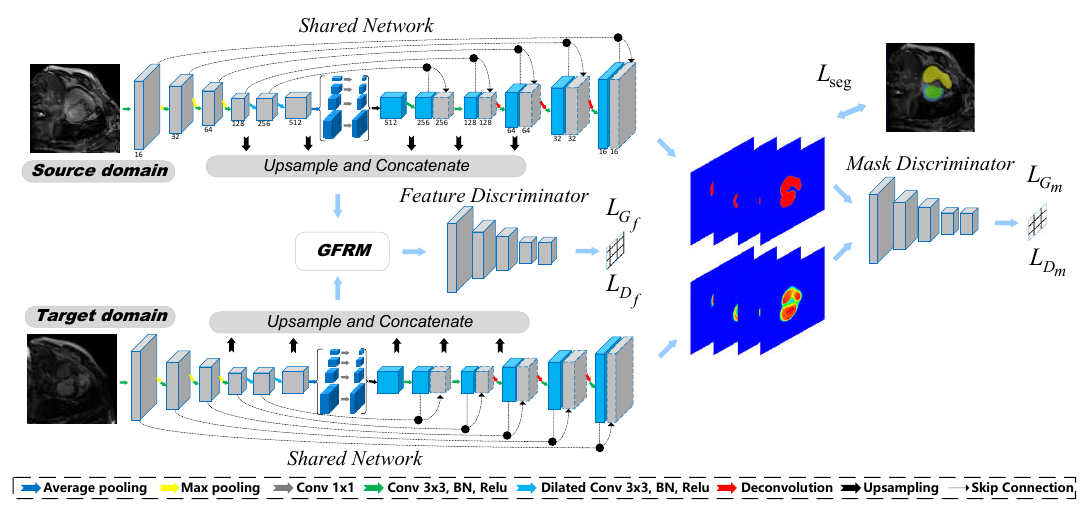}
   \caption{Overview of the multi-modal image translation network proposed by XMU. }
\label{fig:method:XMU}\end{figure}

Particularly, as \zxhreffig{fig:method:XMU} shows, images from the source domain and target domain are fed into a shared modified U-Net for segmentation. The output is the corresponding segmentation.
Within the segmentation network, feature maps of different sizes are combined by upsampling and concatenating, which are then sent to the group-wise feature re-calibration module (GFRM), where a feature-level discriminator measures the discrepancy in the feature space.
The predicted masks from source and target domains are also compared by a mask-level discriminator as an additional constraint.

\subsection{INRIA Sophia Antipolis (INRIA)} \label{method:INRIA}

INRIA proposes to use style data augmentation to prevent the model from over-fitting to any specific contrast and to focus the optimization on the fundamental geometry features of the target \citep{conf/STACOM/ly2020}.
To increase the diversity of training images from the source domain (T2 and bSSFP), multiple image processing functions are applied to the source images, including adaptive histogram equalization, Laplacian transformation, Sobel edge detection, intensity inversion and histogram matching towards the target modality (LGE).

The dual U-Net strategy is adopted \citep{conf/STACOM/jia2018}, where the two networks are trained independently.
These two networks, consecutively applied in the segmentation pipeline, are responsible for localization of the epicardium and refinement of the segmentation, respectively.
Additional improvement on the segmentation accuracy is attained with the proposed thresholded connection
layer network (TCL-Net), which is able to eliminate non-target pixels and to highlight the target pixels of the input image.

\subsection{Sichuan University (SCU)} \label{method:SCU}

SCU uses a supervised method and is based on a neural network primarily modified from a 2D U-Net,
where an input sample consists of three channel 2D images extracted from three neighbored slices in an image.
The segmentation network is equipped with the squeeze-and-excitation residual (SE-Res) module and the selective kernel (SK) module, inserted in the encoding and decoding stages, respectively \citep{conf/STACOM/wang2020}.
The SE-Res module adaptively recalibrates channel-wise feature responses by explicitly modelling interdependencies between channels \citep{conf/cvpr/hu2018}, while the SK module is supposed to adaptively adjust the size of its receptive field based on multiple scales of input information \citep{conf/cvpr/li2019}.
A series techniques, including hole filling and connected component analysis, are employed to post-process the output of the segmentation network.


\subsection{University of Barcelona~(UB)}  \label{method:UB}

\begin{figure}[t]\center
 \includegraphics[width=0.48\textwidth]{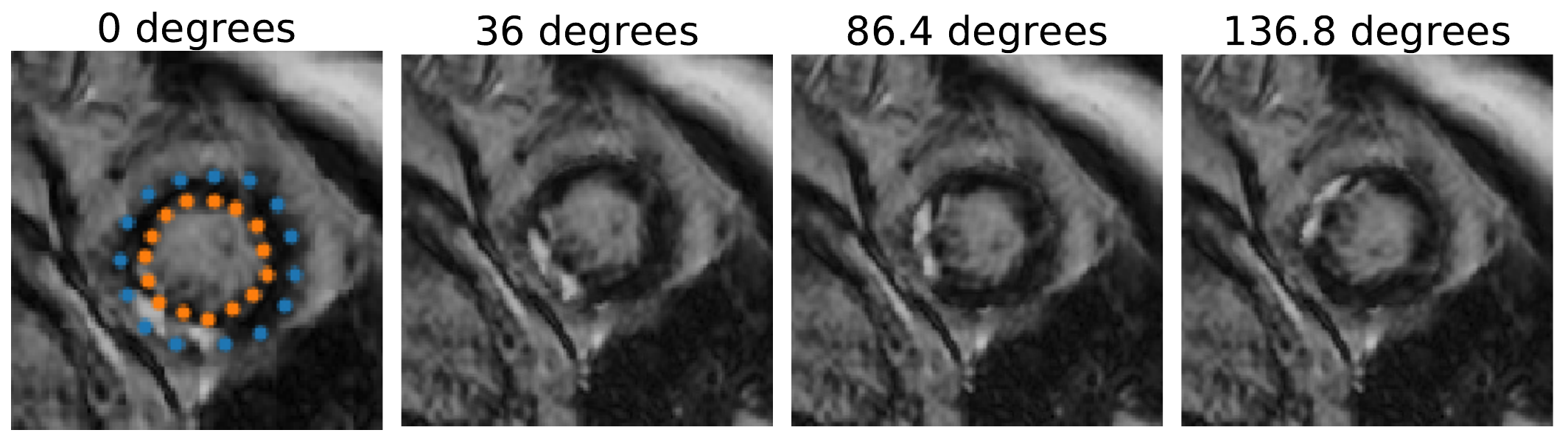}
   \caption{LGE-specific data argumentation strategy: examples of three rotations of myocardium by landmarks shown in the first column, proposed by UB.
   }
\label{fig:method:UB}\end{figure}

UB proposes to address the limited availability of training samples for LGE CMR segmentation by enriching the CNN models using two complimentary methods \citep{conf/STACOM/campello2019}.
First, an image-to-image translation generative adversarial network is trained to convert bSSFP images into LGE-like ones using the CycleGAN strategy \citep{conf/ICCV/zhu2017}.
Second, an LGE-specific data argumentation scheme, i.e., partially region rotation of scars, is proposed based on the provided labeled LGE images, as shown in \zxhreffig{fig:method:UB}.
The image synthesis and scar tissue augmentation strategies can account for both the global appearance of LGE image and the local appearance of scar tissues.
Based on the original sequences and synthetic LGE images from image translation, a modified U-Net network is then trained for LGE segmentation by integrating a deep supervision term in the upsampling path of the U-Net.

\subsection{Friedrich-Alexander-Universit\"at Erlangen-N\"urnberg~~~~ \\(FAU)} \label{method:FAU}
FAU employs a transfer learning strategy, by training an encoder-decoder segmentation network using labeled T2 and bSSFP images, and then fine-tuning the network with the labeled LGE images for supervised training \citep{conf/STACOM/vesal2019}.
The encoder-decoder network, called Dilated-Residual U-Net \citep{conf/STACOM/vesal2018}, can capture both global and local context information using dilated convolutional layers in the bottleneck.
Moreover, before the images are fed to the network, the image contrast is enhanced slice-by-slice, using contrast limited adaptive histogram equalization.
Besides, connected component analysis as a post-processing step is employed to separately remove the small misclassified areas in the softmax layer output for Myo, LV, and RV.

\subsection{NVIDIA} \label{method:NVIDIA}

NVIDIA combines classical methods of multi-atlas label fusion with deep learning by creating the so-called noisy labels for unlabeled LGE images using registration techniques \citep{conf/STACOM/roth2019}.

The noisy labels are derived from two sets of registrations.
One is the inter-patient and intra-modality registration from the annotated 5 LGE images to the remaining 40 ones.
The other is the intra-patient and inter-modality registration from the corresponding annotated bSSFP/ T2 to the 40 unlabeled LGE images.
The registration process contains affine transformation followed by free-form deformation in a coarse-to-fine manner, where normalized mutual information is employed as the similarity measure.
For deformation regularization, the bending energy on the deformation fields is used to encourage smoothness and a penalty term to encourage inverse consistency is also included.

Multi-atlas label fusion is conducted before the fused pseudo labels are used to supervise an ensemble of segmentation networks.
The networks are constructed using convolutional kernels from the pretrained \emph{ImageNet} weights.
Besides, the weights are initialized in three orthogonal planes to encourage view discrepancy.

The final prediction is achieved by fusing probability maps from each model using a median operator followed by a post-processing step of 3D connected component analysis to reduce outliers.

\begin{table*} [t] \center
\caption{
Results of the evaluated algorithms on LGE CMR segmentation.
Teams updating their results after the challenge deadline are indicated with an asterisk (*), and teams
using the unlabeled LGE images ($I^\mathrm{unl}$) for training are indicated with a dag ($\dag$). 
}
\label{tb:result:DiceHD}
{\footnotesize
\begin{tabular}{ |l|lll|lll|l|}\hline
\multirow{2}*{Teams} & \multicolumn{3}{c|}{Volumetric Dice} & \multicolumn{3}{|c|}{Volumetric HD (mm)} & \multirow{2}*{Training}\\
\cline{2-7}
~ &  Myo & LV & RV &   LV Endo &   LV Epi &  RV Endo & ~\\
\hline

ICL$^\dag$&  $0.826 \pm 0.035$	& $0.919 \pm 0.026$ & $0.875 \pm 0.050$ & $10.28 \pm 3.376$	& $12.45 \pm 3.142$	& $15.38 \pm 6.942$ & Unsupervised\\
XMU$^\dag$&  $0.796 \pm 0.059$	& $0.896 \pm 0.047$ & $0.846 \pm 0.086$ & $13.59 \pm 5.206$	& $15.70 \pm 5.814$	& $15.21 \pm 6.327$ & Unsupervised\\
INRIA$^*$ &  $0.705 \pm 0.115$	& $0.870 \pm 0.051$ & $0.762 \pm 0.150$ & $41.74 \pm 7.696$	& $42.79 \pm 13.26$	& $34.38 \pm 8.065$ & Unsupervised\\
\hline
SCU$^*$   &  $0.843 \pm 0.048$	& $0.926 \pm 0.028$ & $0.890 \pm 0.044$ & $9.748 \pm 3.280$	& $11.65 \pm 4.002$	& $13.34 \pm 4.615$ & Supervised \\
UB$^\dag$&  $0.810 \pm 0.061$	& $0.898 \pm 0.045$ & $0.866 \pm 0.050$ & $10.78 \pm 4.066$	& $11.96 \pm 3.620$	& $15.91 \pm 6.895$ &  Supervised\\
FAU	   &  $0.789 \pm 0.073$	& $0.912 \pm 0.034$ & $0.833 \pm 0.084$ & $11.29 \pm 4.559$	& $12.54 \pm 3.379$	& $17.11 \pm 6.141$ &  Supervised\\
NVIDIA$^\dag$&  $0.780 \pm 0.047$	& $0.890 \pm 0.043$ & $0.844 \pm 0.063$ & $11.58 \pm 7.524$	& $16.25 \pm 6.336$	& $18.12 \pm 9.262$ &  Supervised\\
HIT$^\dag$&  $0.751 \pm 0.119$	& $0.884 \pm 0.070$ & $0.791 \pm 0.165$ & $14.30 \pm 8.170$	& $14.75 \pm 7.823$	& $17.87 \pm 9.322$ &  Supervised\\
SUSTech	  &  $0.610 \pm 0.102$	& $0.824 \pm 0.068$ & $0.710 \pm 0.135$ & $23.69 \pm 14.66$	& $24.62 \pm 12.66$	& $23.46 \pm 7.596$ &  Supervised\\
\hline \hline
\multirow{3}*{Average} &  $0.775 \pm 0.093$ & $0.895 \pm 0.047$ & $0.828 \pm 0.114$ & $21.87 \pm 15.23$ & $23.65 \pm 16.07$ & $21.66 \pm 11.49$ & Unsupervised\\
~		 &  $0.764 \pm 0.109$ & $0.889 \pm 0.060$ & $0.822 \pm 0.117$ & $13.56 \pm 9.316$ & $15.30 \pm 8.389$ & $17.64 \pm 8.092$ &  Supervised\\
~		 &  $0.766 \pm 0.104$ & $0.891 \pm 0.056$ & $0.822 \pm 0.116$ & $16.37 \pm 12.27$ & $18.06 \pm 12.18$ & $19.35 \pm 9.587$ & All\\
\hline \hline
Inter-Ob   & $ 0.764 \pm 0.069 $ & $ 0.881 \pm 0.064 $&  $ 0.816 \pm 0.084 $ & $ 12.03 \pm 4.443 $ & $ 14.32 \pm 5.164 $&  $ 21.53 \pm 9.460 $&Inter-Ob\\
\hline
\end{tabular} }\\
\end{table*}

\begin{table*} [t] \center
	\caption{
	Details on the average run time and computer systems used for the evaluated methods.
	T: average run time in seconds (including pre- and post-processing of images).}
\label{tb:result:computer_time}
{\small
\begin{tabular}{ l| l l l l *{4}{@{\ \,}  l}}\hline
Teams	   &  T(s)  &  GPU  & CPU and RAM  & Platform  \\
\hline

ICL	 &   0.92 & GeForce RTX 2080Ti; 11GB  & Intel Core i9-9900K; 64GB (segmentation)  & Pytorch \\ 
			&   &  & Intel Core i7 8700; 32GB (translation) &  \\
XMU		 &15  &GeForce GTX 1080Ti; 12GB & Intel i5-7640X; 16GB	 & Pytorch\\ 
INRIA	   &   0.958	& GeForce GTX 1080Ti; 12GB & Intel Xeon Silver 4110; 8GB & Keras \\ 
\hline
SCU		 & 12   &Tesla P40; 24GB   & Intel(R) Xeon(R) E5-2680; 128GB  & Pytorch\\  
UB	  &   6.3	& GeForce 1080; 8GB & Intel Core i7-8700X; 32GB & Tensorflow/ Pytorch   \\ 
FAU	   &2.5   & Titan X-Pascal; 12GB; & Intel(R) Xeon(R) 4114; 32GB  & TensorFlow/ Keras \\
NVIDIA		& 6.64   &Tesla V100; 16GB (training)   & Intel(R) Core(TM) i7-7800X; 32GB & TensorFlow\\ 
					&  &Titan Xp; 12GB (testing)   & & \\
HIT		&  10   & GeForce RTX 2080Ti; 12GB  & Intel(R) Xeon(R) Silver 4114; 32GB   &Keras \\
SUSTech	 &  60  &  GTX Titan X; 12GB		 &  Intel Core i7-6700; 64GB	& TensorFlow/ Keras \\

\hline
\end{tabular} }\\
\end{table*}

\subsection{Harbin Institute of Technology (HIT)} \label{method:HIT}
HIT adopts an intensity histogram match technology to achieve data augmentation.
The method is developed as follows:
First, the provided bSSFP images, with gold standard segmentation, are intensity-translated to generate fake LGE CMR images, by referring to the real LGE CMR and using histogram matching techniques.
The resulting LGE images have the same shape of the heart, as the original bSSFP data, and thus the gold standard segmentation of the bSSFP can be used for the fake LGE images.
Then, the real labeled LGE CMR images and the fake LGE with gold standard are used to train a Res-UNet in the supervised learning fashion \citep{conf/CVPR/he2016}.
To improve the segmentation accuracy, the output of the segmentation network is further post-processed using the label-vote strategy and connected component analysis \citep{conf/STACOM/liu2019}.

\subsection{Southern University of Science and Technology \\(SUSTech)}\label{method:UD}
SUSTech proposes an adversarial segmentation framework, where a discriminator is used to drive the generator to produce good-quality masks similar to the ground truth segmentation \citep{conf/STACOM/chen2019}.
Specifically, the method transfers the label masks of bSSFP and T2 CMR images to LGE CMR by using a normalized index which identifies the correspondence between axial slices from different modalities.
Then, the generator model, formed by a segmentation network, is trained to output realistic LGE labels with the combination of supervised segmentation loss and discriminator loss.

\begin{figure*}[t]\center
	\includegraphics[width=1\textwidth]{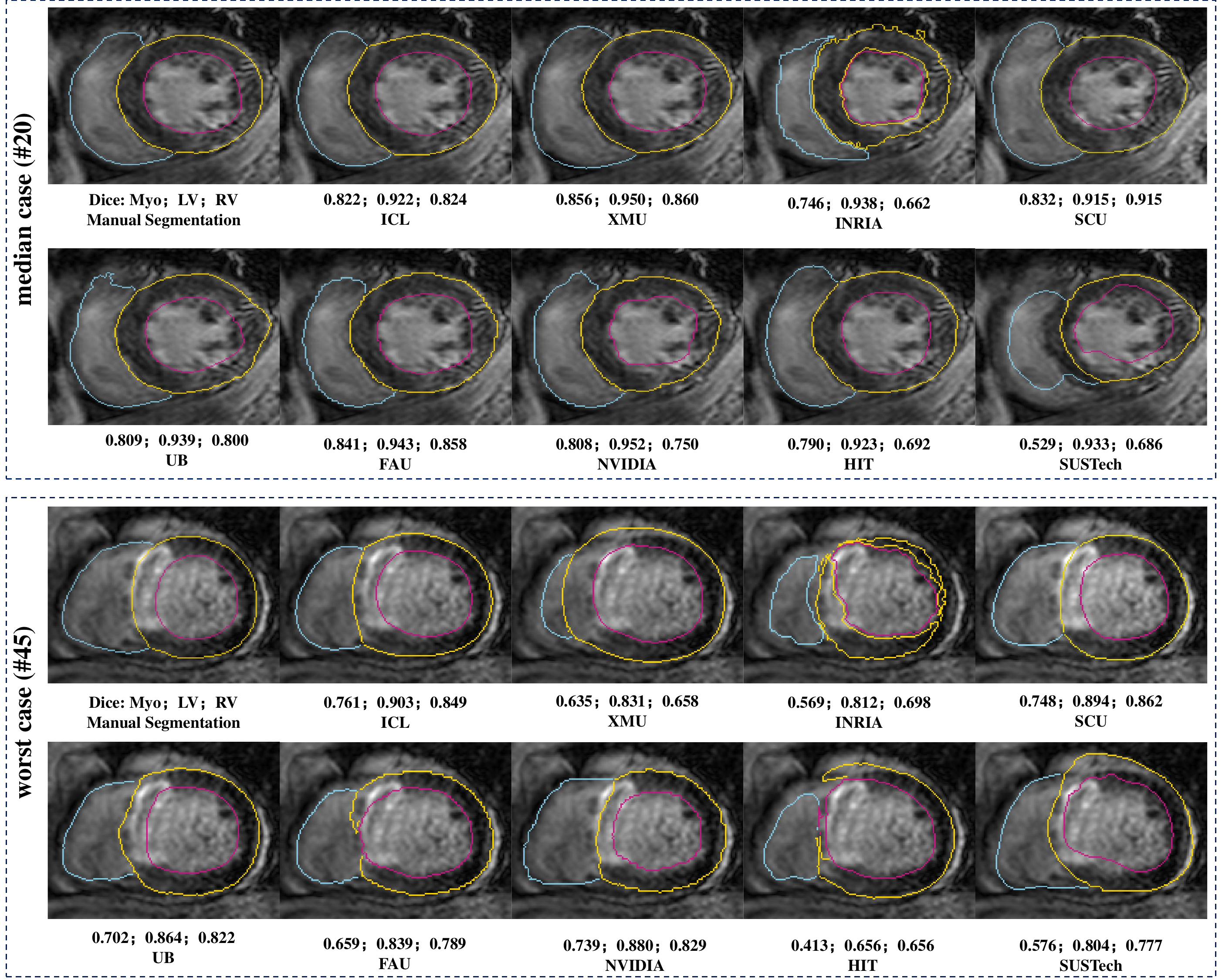}\\[-1ex]
	\caption{2D visualization of the segmentation results of the median and worst cases by the nine evaluated methods.
	The median and worst cases were from the test set in terms of mean Dice of Myo by the nine methods.
		The contours of LV Endo, LV Epi and RV are respectively in colors of magenta, yellow and blue.
	}
\label{fig:exampleseg}\end{figure*}

\begin{table*}[th] \center
\caption{
Slice-wise accuracies of the apical (Apex), mid-ventricular (Mid), and basal (Base) slices.
}
\label{tb:slice:DiceHD}
{\footnotesize
\begin{tabular}{ |l|lll|lll|l|}\hline
\multirow{2}*{Teams} & \multicolumn{3}{c|}{Slice-wise Dice} & \multicolumn{3}{|c|}{Slice-wise HD (mm)} & \multirow{2}*{Training}\\
\cline{2-7}
~ &  Apex & Middle & Base &   Apex &   Middle &  Base & ~\\
\hline
ICL$^\dag$&  $0.706 \pm 0.203$	&	$0.883 \pm 0.078$	&	$0.857 \pm 0.140$	&	$11.24 \pm 10.73$	&	$6.140 \pm 5.144$	&	$8.229 \pm 7.306$ & Unsupervised\\
XMU$^\dag$&  $0.502 \pm 0.292$	&	$0.844 \pm 0.136$	&	$0.828 \pm 0.165$	&	$20.76 \pm 17.64$	&	$7.604 \pm 8.775$	&	$8.992 \pm 7.758$ & Unsupervised\\
INRIA$^*$ &  $0.612 \pm 0.230$	&	$0.795 \pm 0.142$	&	$0.724 \pm 0.221$	&	$14.45 \pm 15.52$	&	$12.74 \pm 13.57$	&	$16.99 \pm 15.93$ & Unsupervised\\
\hline
SCU$^*$ &  $0.705 \pm 0.203$	& $0.895 \pm 0.076$ & $0.894 \pm 0.072$ & $10.12 \pm 9.639$ & $5.300 \pm 4.823$ & $6.304 \pm 3.875$ & Supervised \\
UB$^\dag$&  $0.694 \pm 0.218$	&	$0.855 \pm 0.103$	&	$0.822 \pm 0.170$	&	$11.45 \pm 11.30$	&	$7.331 \pm 6.670$	&	$10.45 \pm 10.68$ &  Supervised\\
FAU	 &  $0.622 \pm 0.268$	&	$0.831 \pm 0.148$	&	$0.864 \pm 0.109$	&	$17.35 \pm 15.87$	&	$9.753 \pm 10.51$	&	$8.572 \pm 6.966$ &  Supervised\\
NVIDIA$^\dag$& $0.727 \pm 0.158$ &	$0.852 \pm 0.085$	&	$0.821 \pm 0.110$	&	$10.64 \pm 8.371$	&	$8.775 \pm 9.029$	&	$11.09 \pm 8.482$ &  Supervised\\
HIT$^\dag$&  $0.582 \pm 0.275$	&	$0.807 \pm 0.179$	&	$0.771 \pm 0.240$	&	$17.81 \pm 15.90$	&	$9.864 \pm 11.35$	&	$14.66 \pm 16.00$ &  Supervised\\
SUSTech	  &  $0.594 \pm 0.202$	&	$0.734 \pm 0.155$	&	$0.736 \pm 0.150$	&	$14.96 \pm 9.106$	&	$12.54 \pm 9.612$	&	$16.30 \pm 11.24$ &  Supervised\\
\hline \hline
\multirow{3}*{Average} &  $0.607 \pm 0.258$	&	$0.841 \pm 0.127$	&	$0.803 \pm 0.187$	&	$15.49 \pm 15.43$	&	$8.829 \pm 10.19$	&	$11.40 \pm 11.75$ & Unsupervised\\
~ &  $0.654 \pm 0.231$	&	$0.829 \pm 0.140$	&	$0.818 \pm 0.161$	&	$13.72 \pm 12.50$	&	$8.927 \pm 9.231$	&	$11.23 \pm 10.81$ &  Supervised\\
~ &  $0.638 \pm 0.242$	&	$0.833 \pm 0.136$	&	$0.813 \pm 0.170$	&	$14.31 \pm 13.57$	&	$8.894 \pm 9.563$	&	$11.29 \pm 11.13$ & All\\
\hline \hline
Inter-Ob	&	$0.789 \pm 0.102$	&	$0.858 \pm 0.074$	&	$0.851 \pm 0.089$	&	$8.975 \pm 7.511$	&	$7.444 \pm 5.076$	&	$14.35 \pm 13.38$ &Inter-Ob \\
\hline
\end{tabular} }\\
\end{table*}

\begin{figure*}[th]\center
	\includegraphics[width=0.98\textwidth]{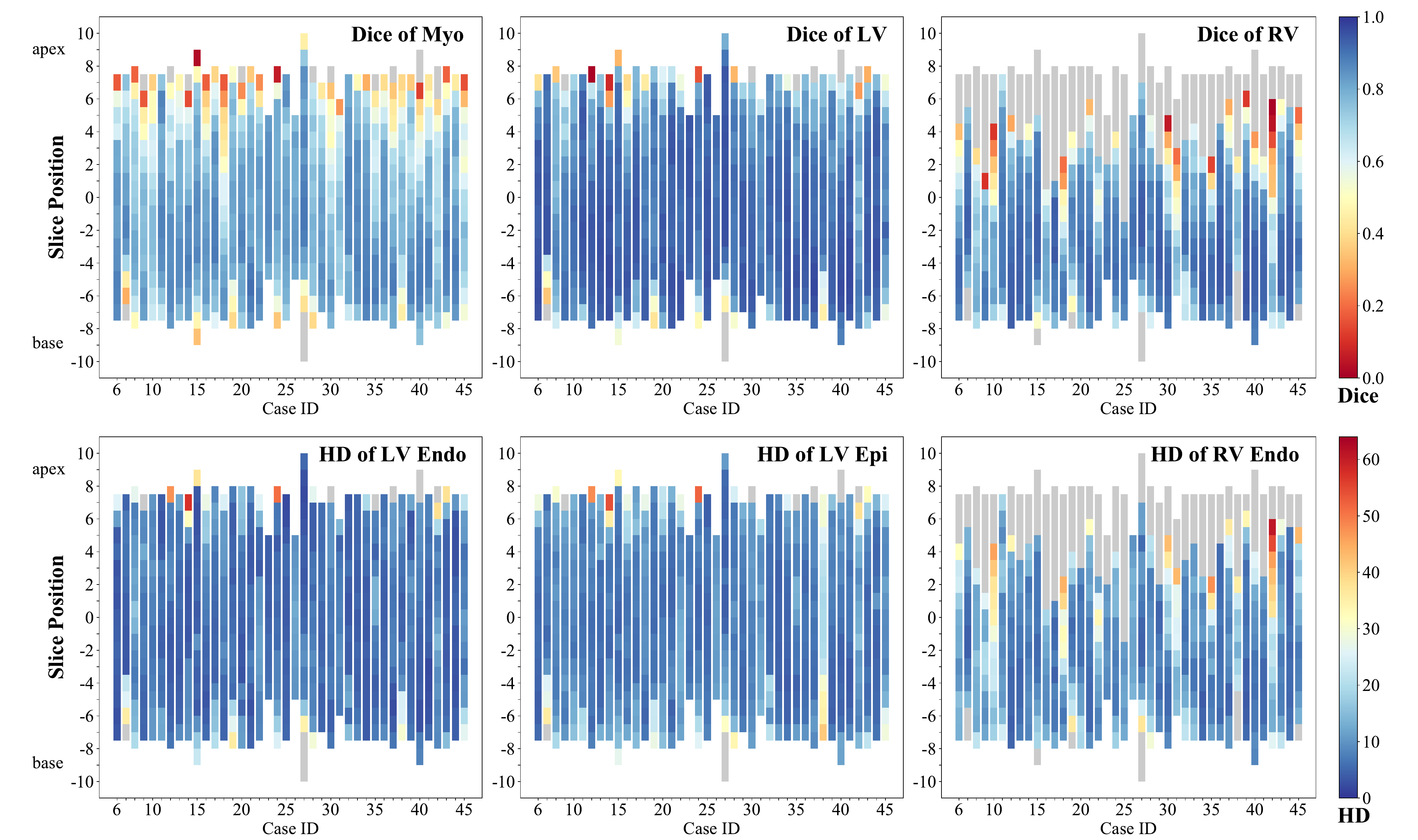}\\[-2ex]
	\caption{The average Dice and HD values of the nine evaluated methods on each slice of the 40 test LGE CMR images.
		Note that grey color indicates that the slices do not contain the manual segmentation of the substructure, and thus no accuracy metric is computed.
	}
	\label{fig:sliceDiceHD}\end{figure*}

\section{Results}\label{result}
\subsection{Performance of the evaluated methods}\label{result:performance}

\Zxhreftb{tb:result:DiceHD} presents the quantitative results of the nine evaluated algorithms on the LGE CMR segmentation,
where the inter-observer variations over the manual delineation are given.
The mean volumetric Dice scores of the evaluated algorithms are respectively $0.766 \pm 0.104$ (Myo), $0.891 \pm 0.056$ (LV), and $0.822 \pm 0.116$ (RV),
and the mean volumetric HD values are respectively $16.37 \pm 12.27$ mm (LV Endo), $18.06 \pm 12.18$ mm (LV Epi), and $19.35 \pm 9.587$ mm (RV Endo).
The mean Dice scores are comparable to that of the inter-observer Dice scores.
Compared with inter-observer HD values, the average HD is however evidently worse, though the top-ranking methods could perform much better.
Interestingly, the unsupervised methods performed comparably to supervised ones, particularly if we exclude INRIA and SUSTech, each respectively from the two categories. 

\zxhreffig{fig:exampleseg} visualizes the segmentation results of case 20 and 45, which are the median (case 20) and worst (case 45) cases in terms of average Dice of the nine methods.
Note that the presented slices are the middle one of these two cases.
Most of the evaluated algorithms could segment case 20 well, though some performed bad on RV segmentation. 
By contrast, case 45 was much more challenging. Over one-fourth of the myocardium in the visualized slice was scarring tissues, which could result in failure for most of the methods due to their similar intensity distribution to the blood pool.
One can see that seven out of the nine methods misclassified a great portion of the scars into blood pool.

\Zxhreftb{tb:result:computer_time} summarizes the information of implementation platforms and average run time of the evaluated methods.
Most of them can complete a segmentation case within seconds on average for those deep neural network-based methods, thanks to the parallel computing of GPU.
Note that the run time includes the pre- and post-processing of the methods, and it may not indicate the true computation complexity of the algorithms due to the different implementation platforms and hardwares.

\subsection{Accuracies of different substructures}  \label{discussion:substructures}

From \Zxhreftb{tb:result:DiceHD}, one can see that the Dice scores of LV are evidently better than that of RV and Myo.
The HD values of myocardium, including LV Endo and Epi, are generally better than that of RV Endo.
This indicates the different levels of challenges for different substuctures, namely RV presents more difficulties.
This could be owing to the fact that LV has a relatively regular shape, in contrast to RV which has large shape variations as well as poor contrast.
This is confirmed by the inter-observer variation study, where the mean HD between different raters could reach over 20 mm and mean Dice only 0.816 for RV, due to the challenge of achieving consistent segmentation on certain areas.
Also, \zxhreffig{fig:exampleseg} confirms that the results of RV segmentation by different methods could vary more than the other substructures.


\begin{figure*}[tp]\center
\includegraphics[width=0.32\textwidth]{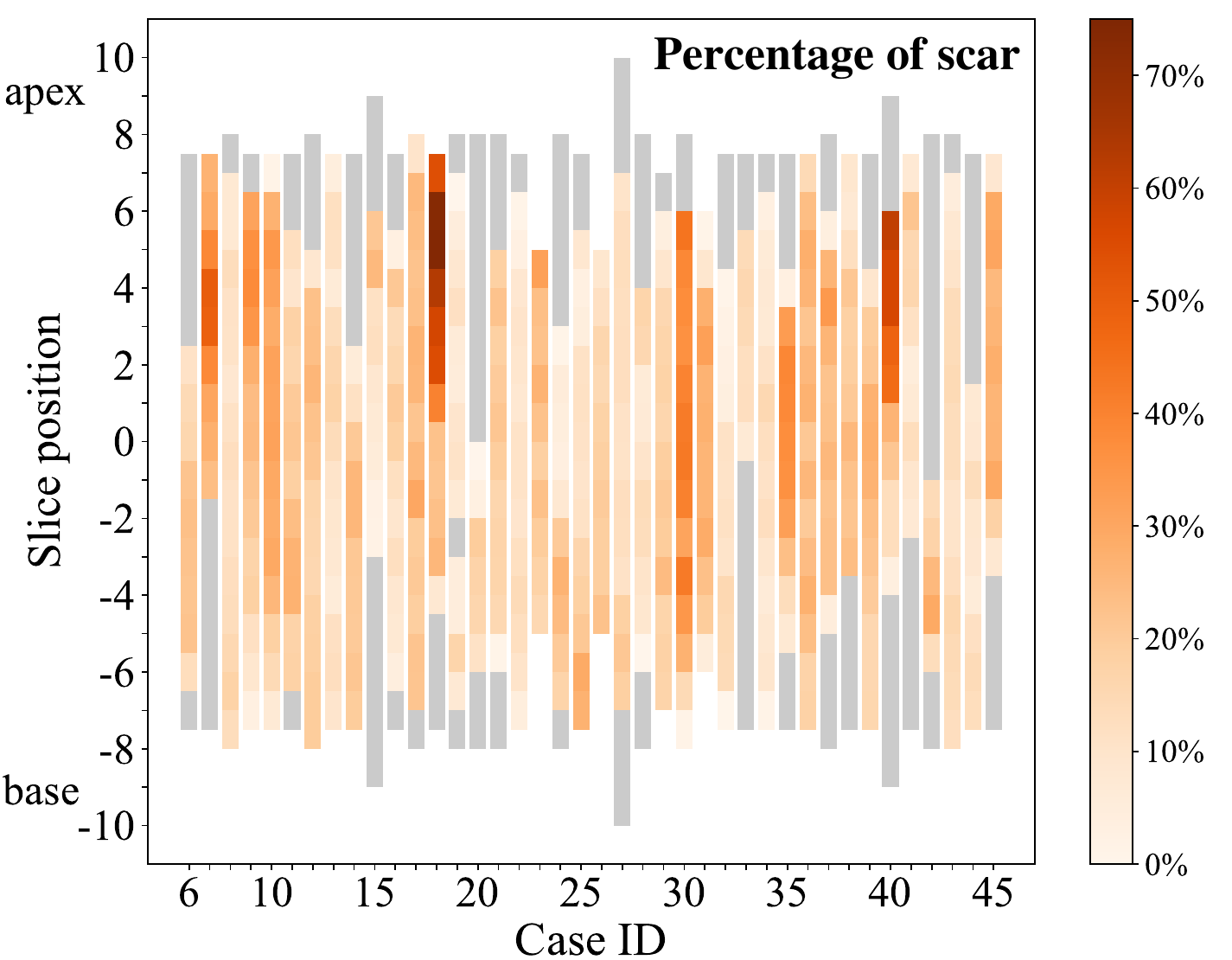}
\includegraphics[width=0.32\textwidth]{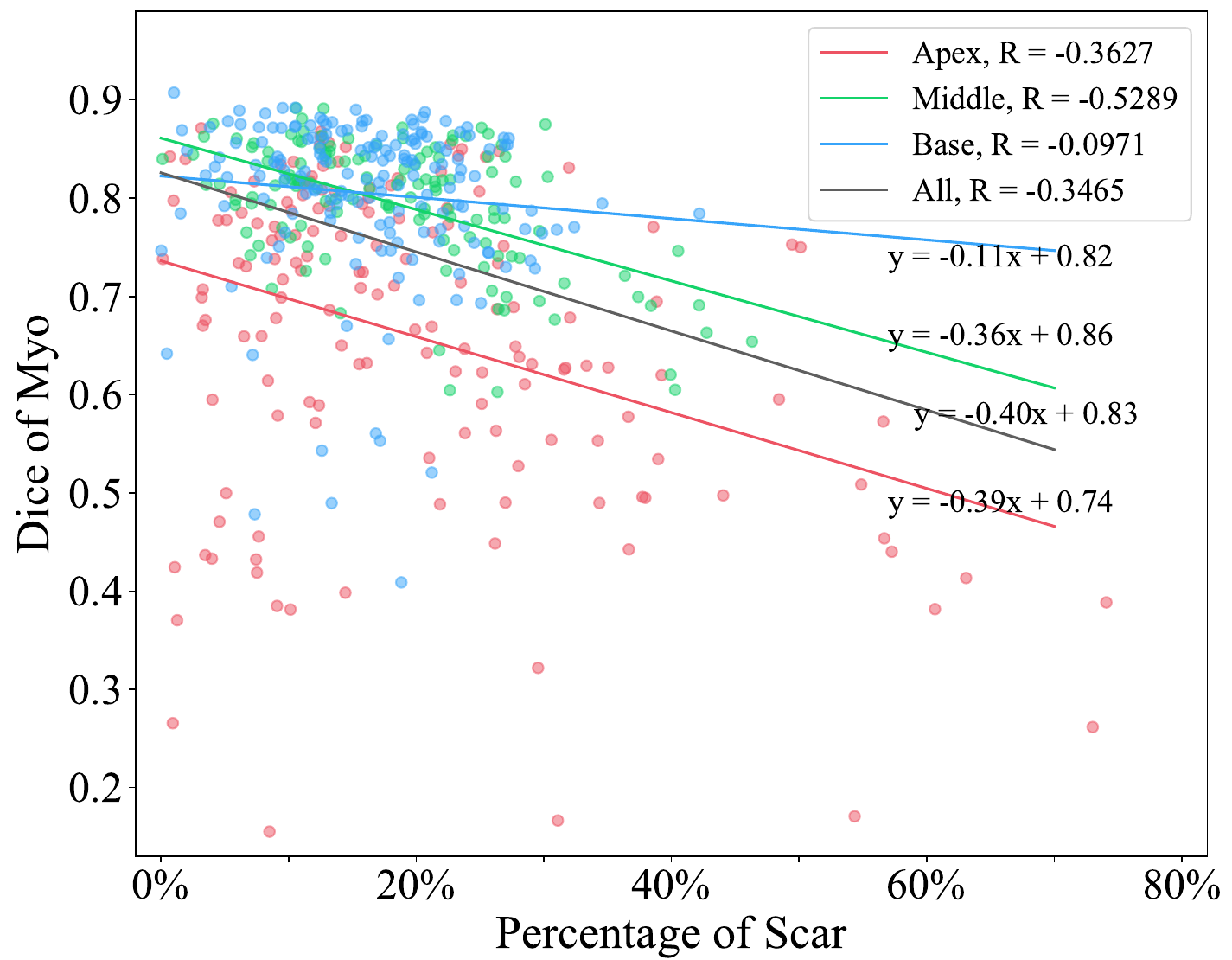}
\includegraphics[width=0.32\textwidth]{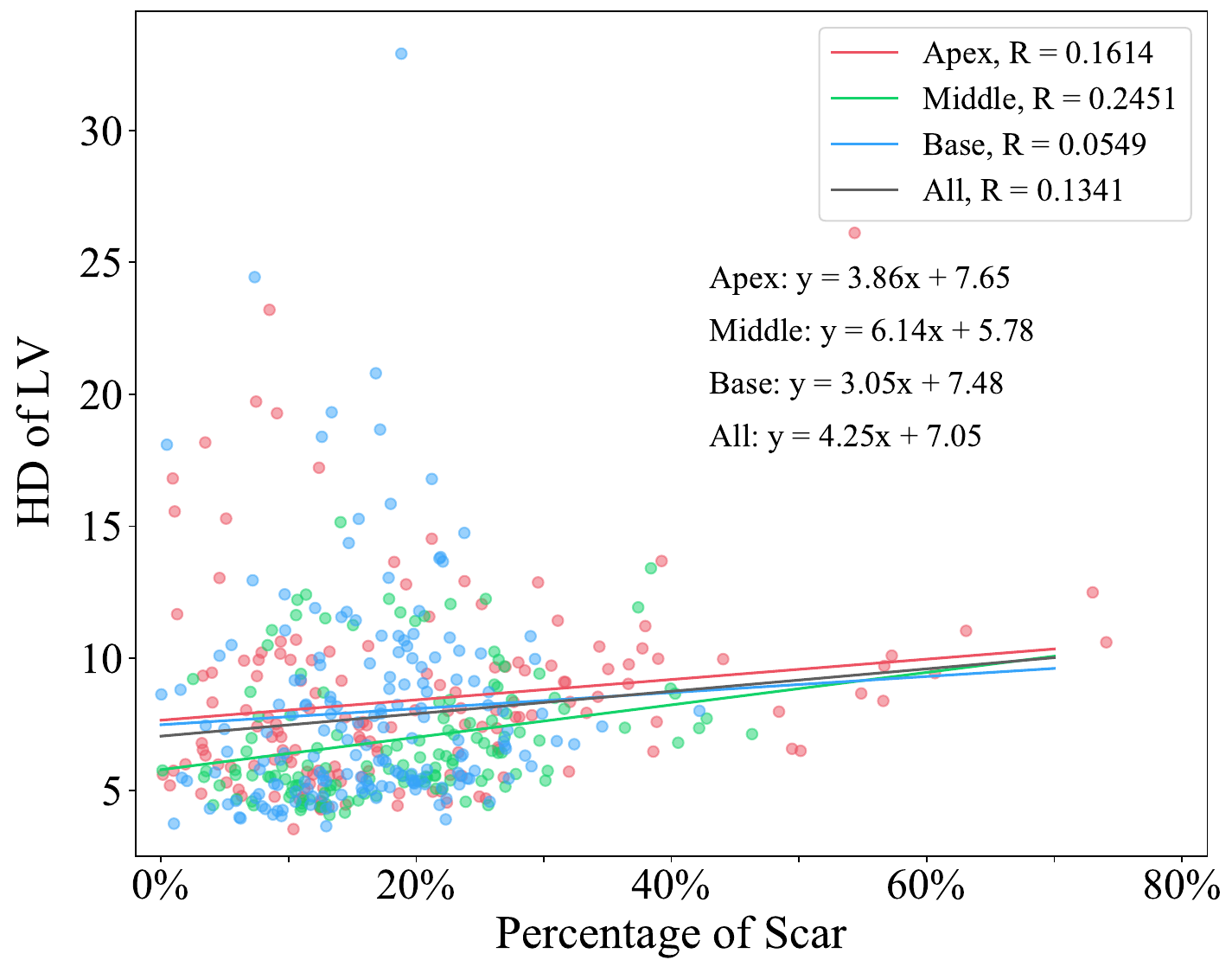}\\
\makebox[\linewidth][s]{ (a)\ \ (b)\ \ (c) }\\[-2ex]
\caption{Results of scar presence study:
(a) The color map of scar percentage (to myocardium) in each slice of the test images, where grey color indicates no scar is presented.
(b) The correlation between the Dice of Myo and the scar percentage for slices from different positions.
(c) The correlation between the HD of LV (mean of HD of LV Endo and Epi) and the scar percentage.
Apex, Mid, and Base respectively represent the apical, mid-ventricular and basal slices, and All indicates all slices are considered, whose points were omitted in (b) and (c).
}
\label{fig:ScarSummary}\end{figure*}

\subsection{Slice-wise accuracies from different positions} \label{result:slice}
\Zxhreftb{tb:slice:DiceHD} presents the slice-wise Dice and HD values for the slices from different positions of the ventricles, including the apical slices (denoted as Apex), mid-ventricular slices (as Middle) and basal slices (as Base). 
Both of the Dice and HD metrics were computed by averaging all the numbers from the three substructures, i.e., Myo, LV, RV for Dice, and LV Endo, LV Epi and RV Endo for HD.
One can see that the different methods could have different superiority/ inferiority regarding segmentation for the slices from different positions. For example, XMU and FAU had particularly poor Dice and HD for apical slices, though they performed well for the Middle and Base; UB achieved superior HD and Dice on Apex and Middle compared with most of the algorithms, but its segmentation on Base was evidently poor.

\Zxhreftb{tb:slice:DiceHD} also provides the inter-observer variations.
The inter-observer Dice and HD values of Apex are evidently poorer than that of Middle and Base; but for HD the worst slice comes from the Base.
Both of the two metrics show good values for middle slices, indicating less difficulties and more reliability for manual delineation, which is consistent with the automatic segmentation.

Compared to the inter-observer variations, none of the evaluated methods could achieve satisfactory accuracies on Apex. In fact, their mean Dice or HD values are all much worse.
This is unlike the results in \Zxhreftb{tb:result:DiceHD} where majority of the evaluated methods could achieve better Dice and HD values than the inter-observer variations.
Such poor performance of the automatic segmentation on Apex is well illustrated in \zxhreffig{fig:sliceDiceHD},
which colorizes the mean accuracies of the three substructures evaluated on the nine methods, for all slices of the test dataset.
One can see from the figure that the apical slices are generally with much worse accuracies, particularly for segmentation of LV Myo and RV.


\subsection{Myocardial segmentation in presence of scars}\label{result:scar}
Besides the inferior image quality of LGE CMR compared to the bSSFP CMR, the inhomogeneous intensity of myocardium with the presence of enhanced scars could particularly challenge an automatic segmentation algorithm.
This was illustrated by Case 45, \zxhreffig{fig:exampleseg}, where the scar area resulted in the failure of most of the evaluated algorithms.
In this study, we investigate the correlation of myocardial segmentation accuracies with the presence and amount of scars in the images.

\zxhreffig{fig:ScarSummary} (a) visualizes the scar percentage for each slice of the test images,
i.e., the area percentage computed as the ratio of scar area to the total area of myocardium in a slice.
One can compare this figure to the visualizations in \zxhreffig{fig:sliceDiceHD}, to study the correlation between slice positions and scar percentage, which nevertheless is not evident due to the complex patterns of scars.
We further plotted the scatter points describing the accuracies of myocardial segmentation against the scar percentage for each slice containing myocardial infarcts,
in \zxhreffig{fig:ScarSummary} (b) and (c).
One can see that the Dice accuracies of myocardial segmentation and scar percentage are generally negatively correlated, though this correlation is very weak for basal slices.
This is reasonable as the scar percentage of basal slices is generally small, and the challenge of segmentation in Base could be mainly attributed to the other factors such as irregular shape of the ventricles.
The HD errors in \zxhreffig{fig:ScarSummary} (c) illustrate positive correlation to scar percentage, which is similar to inverse Dice metric.
However, the correlations of HD are generally weaker than that of Dice.

\section{Discussion} \label{discussion}

\subsection{Supervised versus unsupervised learning}
An interesting observation from this challenge is that the average performance of the supervised methods was not evidently better than that of the unsupervised ones, particularly when the outlier approaches were excluded for comparisons.
One may argue that this was due to the fact that ICL and XMU, the two well-performing unsuperivsed approaches, employed the 40 unlabeled test images as target images in domain adaptation, for the training of their target segmentation networks.
By contrast, INRIA solely used the 5 unlabeled LGE CMR images from the training set, and its accuracies were much poorer.
However, three of the six supervised methods, i.e., UB, NVIDIA and HIT, also used the unlabeled test images for self-learning,
and one can see from \zxhreftb{tb:result:DiceHD} that their average segmentation accuracies were not better than that of the other three (SCU, FAU and SUSTech), nor better than ICL or XMU. 

The other reason could be attributed to the fact that the image appearance of LGE CMR and bSSFP CMR is similar,
and thus the domain shift between them could be small for the unsupervised domain adaptation methods when they considered the LGE CMR as target domain and used the labeled bSSFP CMR as the source domain.
Particularly, we realized that for the supervised methods their models were all first trained by bSSFP (and T2) CMR and then fine-tuned using the 3 to 5 labeled LGE CMR images.
Considering the small domain shift between the two sequences, we conclude that the major performance difference between the supervised and unsupervised approaches could be mainly caused by the 3-5 extra labeled training data on top of the 20-35 ones, more than by the difference of methodologies.

\subsection{Segmentation/ training with multi-sequence CMR} \label{dicsussion: multi-sequences}

Previous discussion indicates the other CMR sequences, particularly the bSSFP CMR, could be critical to the success of LGE CMR segmentation.
For the usage of mult-sequence CMR, three schemes were adopted:
(1) propagate the labels from the other sequences to the target LGE CMR;
(2) directly train the segmentation models using the labeled (or/ and unlabeled) data from the other sequences;
(3) make use the complementary information of the paired multi-sequence CMR of the same subject for segmentation of LGE images.

Among the evaluated methods, only NVIDIA adopted scheme (1), SCU and FAU used scheme (2), and remained teams employed scheme (3).
Specifically, NVIDIA utilized registration-based methods to propagate the label of other sequences to the LGE CMR.
Though the registration between LGE and other sequences can be challenging, it is effective to obtain results with realistic shapes. The readers are referred to the segmentation results of Case 45 in \zxhreffig{fig:exampleseg}, where NVIDIA maintained a reasonably accurate result with realistic shape for this particular challenging case.
SCU and FAU directly trained their segmentation networks using all the labeled data from three sequences.
HIT utilized bSSFP CMR labeled images to train a localization network, which can offer an ROI as prior knowledge for LGE segmentation.
XMU, INRIA and HIT adopted histogram matching, by mapping intensity of bSSFP and T2 CMR images into that of LGE CMR, to generate fake LGE images for supervised training of their models.
ICL and UB used a style transfer network to synthesize LGE images, and XMU and SUSTech generated pseudo-LGE labels from other sequences.
One can see that the paired corresponding information among different sequences can be learned implicitly by directly adopting other sequences of the same patient as extra training data.
To generate pseudo-LGE images/ labels, style transfer demonstrated to be more effective than the conventional methods such as histogram matching.

\subsection{Discussion of the literature}

In the preceding discussion, we summarized the three usages of mult-sequence CMR for LGE image segmentation.
In the literature, the development of automated segmentation on LGE CMR follows the similar three schemes:
\begin{enumerate}
\item[(1)] Propagate the prior labels from another sequence (typically bSSFP CMR) from the same patient to the target LGE CMR \citep{journal/mia/Wei2013,journal/jcmr/Lu2013,journal/jmri/Tao2015}.
\item[(2)] Directly apply the segmentation method on the LGE CMR images
\citep{journal/TBME/liu2017,journal/CARS/liu2018,conf/MICCAI/yue2019}. Images from other CMR sequences, commonly not paired images from the same subject of the target LGE CMR, could be used to assist the segmentation or training of the neural networks.
\item[(3)] Combine the complementary information of the paired multi-sequence CMR of the same subject for segmentation of LGE CMR \citep{conf/ISBI/liu2014,journal/cmig/liu2019,journal/pami/Zhuang2019}.
\end{enumerate}
\Zxhreftb{tb:discuss:literature} summarizes the works of these three categories from recent literature.

One can see that most of the reported works adopted multi-sequence CMR for the segmentation of LGE CMR, except the two works from \citet{journal/TBME/liu2017,conf/MICCAI/yue2019}.
For the auxiliary sequences, images from bSSFP CMR are generally useful.
This is because bSSFP images have similar appearance to LGE. This is particularly useful to the first category of methods, where the prior labels are propagated to LGE CMR via image registration.
Another reason is the wide availability of the bSSFP images, which can be used for pre-training or for shape regularization of a deep neural network-based algorithm.
Finally, for the combined segmentation using the paired images from the same patient, both the T2 and bSSFP could be employed. For this category of methods, the registration between LGE and bSSFP/ T2 needs to be simultaneously implemented with the procedure of segmentation, to maintain a consistent segmentation among all sequences of the same patient \citep{journal/pami/Zhuang2019}.

For accuracy, all the reported works of the prior label propagation methods achieved over 0.80 Dice accuracy on myocardium segmentation. For the other methods, only the recent work by \citep{conf/MICCAI/yue2019} obtained over 0.80 Dice of Myo.
The label propagation methods require prior segmentation of the bSSFP images, which are generally manually performed. Hence, they are not considered as fully automatic schemes.
Furthermore, considering the ratio of the Myo Dice to the Myo inter-observer Dice, one can see that neither of the two reported label propagation methods achieved a ratio over 1.
By contrast, the ratio of \citet{conf/MICCAI/yue2019}, i.e., 0.812 versus 0.757, is much better, and is comparable to the top-ranking methods in this benchmark study.
Finally, the combined segmentation incorporating multi-sequence CMR reported relatively lower accuracies.
However, these methods did not require any manual segmentation or labeled CMR images for training \citep{journal/pami/Zhuang2019}.
In the future, we expect more research on the topic of the combined segmentation embedded with deep neural networks for automatic segmentation of LGE CMR.

\begin{table*} [!t] \centering
\caption{
Results of \textit{LGE} CMR segmentation from the literature.
InterOb: inter-observer variations; min: minutes; sec: seconds.
	}
	\label{tb:discuss:literature}
{\small
\begin{tabular}{p{1.5cm}<{\raggedright} p{2.0cm}<{\raggedright} p{4.2cm}<{\centering\raggedright} l   p{7cm}<{\raggedright}  }\hline
Reference	   & Data &  Methodologies & Runtime & Results of LGE seg in Dice and HD (mm) \\
\hline
\cite{journal/mia/Wei2013}	 & 21 (LGE+ bSSFP)  &
Propagate prior segmentation in bSSFP to LGE via 2D translation&N/A&
Myo: $0.828 \pm 0.025$ \qquad\qquad\qquad\qquad\qquad \qquad
LV: $0.948 \pm 0.009$ \qquad\qquad\qquad\qquad\qquad \qquad
InterOb-Myo: $0.835 \pm 0.025$ \qquad\qquad\qquad\qquad
InterOb-LV: $0.951 \pm 0.009$
\\
\cite{journal/jcmr/Lu2013}	   & 10 (LGE+ bSSFP) &
Propagate segmentation in bSSFP to LGE via deformable registration&N/A&
Myo: $0.85 \pm 0.05$ \qquad\qquad\qquad\qquad\qquad\qquad\qquad
LV: $0.95 \pm 0.02$
\\
\cite{journal/jmri/Tao2015}	  & 50 (LGE+ bSSFP)	  &
Propagate segmentation in bSSFP  to LGE via contour fitting refined with scar pattern&N/A&
Myo: $0.805\pm 0.083$ \qquad\qquad\qquad\qquad\qquad\qquad
InterOb-Myo: $0.81 \pm 0.06$
\\\hline

\cite{journal/TBME/liu2017}	  & 33 LGE			&
Coupled level set and multi-component Gaussian mixture model & 7.29 min  &
Myo: $0.736 \pm 0.056$ \qquad\qquad\qquad\qquad\qquad\qquad
LV: $0.905 \pm 0.032$ \qquad\qquad\qquad\qquad\qquad
InterOb-Myo: $0.731 \pm 0.081$ \qquad\qquad\qquad\qquad\qquad
InterOb-LV: $0.886 \pm 0.051$
\\
\cite{journal/CARS/liu2018} & 30 LGE, 51~bSSFP &
Guided random walks and sparse shape representation from manual segmentation of bSSFP&9 min&
Myo: $0.746\pm0.078$\qquad\qquad\qquad\qquad\qquad\qquad\qquad
LV: $0.851\pm0.058$\qquad\qquad\qquad\qquad\qquad\qquad\qquad
InterOb-Myo: $0.739\pm0.051$\\
\cite{conf/MICCAI/yue2019}	   & 45 LGE			&
Deep neural network with shape reconstruction trained from segmentation labels and spatial constraints & N/A   &
Myo: $0.812 \pm 0.105$, Epi: \textit{$11.04 \pm 5.818$} mm \qquad
LV: $0.915 \pm 0.052$, \textit{$12.25 \pm 6.455$} mm \qquad\qquad
RV: $0.882 \pm 0.084$, \textit{$18.07 \pm 14.17$} mm
InterOb-Myo: $0.757 \pm 0.083$
\\ \hline

\cite{conf/ISBI/liu2014}		 & 6 (LGE+T2)  &
Combined segmentation of LGE and T2 based on multi-component Gaussian mixture model &N/A&
Myo: $0.643\pm 0.084$
\\
\cite{journal/cmig/liu2019}	  & 32~(LGE+T2)		&
Combined segmentation of LGE and T2 with cross constraint and discrepancy compensation of shape &15.2 sec&
Myo: $ 0.781 \pm 0.062 $\qquad\qquad\qquad\qquad\qquad\qquad\qquad\qquad
LV: $0.879 \pm 0.044$
\\
\cite{journal/pami/Zhuang2019}   & 35 (LGE+ T2+bSSFP)&
Unified segmentation of three-sequence CMR based on MvMM &9.17 min &
Myo: $ 0.717 \pm 0.076$, Epi: $11.2 \pm 4.05$ mm \qquad\qquad\qquad\qquad
LV: $0.866 \pm 0.063$, $10.6 \pm 4.67$ mm  \qquad\qquad\qquad\qquad
InterOb-Myo: $0.757\pm0.083$, Epi: $12.5\pm5.83$~mm
InterOb-LV: $0.876\pm0.069$, $10.6\pm4.65$ mm \\
\hline
\end{tabular} }
\end{table*}

\section{Conclusion}\label{conclusion}

Cardiac segmentation from LGE CMR is important in clinics for diagnosis and treatment management of patients.
However, fully automatic segmentation is still arduous.
Particularly, the training data with manual labels are limited in the research community of medical image analysis.
This manuscript describes the results from the MS-CMR segmentation challenge, which provides 45 sets of MS-CMR images. Nine representative methods were selected for evaluation and comparisons, and their methodologies and segmentation performance were then analyzed and discussed.

The challenge originally only provided five labeled LGE CMR images for training and validation, and forty for test, to resemble the reality that only very limited labeled training images of LGE CMR are available.
In addition, the challenge provided the paired bSSFP and T2 CMR images for all the LGE CMR images, and a much larger number of bSSFP and T2 images were with gold standard labels for training. The evaluation was performed by the organizers, blinded to the participants for a fair comparison in the challenge.
Note that the gold standard segmentation of the test data as well as the training data had been open to researchers after the workshop of the challenge.

As a result, one third of the evaluated methods adopted the domain adaptation strategy, meaning none of the target domain images, i.e., LGE CMR, has label information in the training stage. The label information came from the source domain, i.e., the bSSFP CMR images with gold standard segmentation.
The other two thirds of methods employed the labeled bSSFP (and T2 images) for pre-training and then fine-tuned the networks with the 3 to 5 LGE CMR images with gold standard.
Interestingly, we did not find evident difference of segmentation performance between these two groups of methods. We concluded this was probably due to the fact that the image appearance of bSSFP CMR and LGE CMR is similar, thus the domain shift is marginal for the domain adaptation-based algorithms.

All the evaluated methods made full use of the MS-CMR images for the development of automatic segmentation of LGE CMR.
This is similar to the development of LGE CMR segmentation from the literature, where most of the reported works used bSSFP and (or) T2 CMR to assist the segmentation of LGE CMR or facilitate the training of a LGE CMR segmentation neural network.
We concluded that the auxiliary sequences from MS-CMR could be helpful and critical for a successful and robust segmentation of LGE CMR.

Finally, several works in the literature employed the paired MS-CMR for combined and simultaneous segmentation of all the CMR sequences, including LGE CMR. However, none of the evaluated methods in this challenge adopted this strategy. We concluded this should be attributed to the fact that consistent segmentation among MS-CMR requires simultaneous registration between sequences in the procedure of combined segmentation, which can be still arduous for deep neural network-based algorithms.
We therefore expect more research on novel methodologies to achieve combined computing with simultaneous registration and segmentation of the multi-source images in the future.

\end{spacing}

\section*{Authors Contributions}
XZ initialized the challenge and provided all the resources;
XZ, LL, JX and XL collected the materials and composed the manuscript.
 CC, CO, DR, VC, KL, SV, NR, YL, GL, JC, HL, BL, MS, HR, WZ, JW, XD, XW, SY
were participants of the MS-CMR segmentation challenge.
The participants provided the description of their algorithms and segmentation results for evaluation,
 and contributed equally to this paper.
All the authors have read and approved the publication of this work.

\section*{Acknowledgement}
This work was supported by the National Natural Science Foundation of China (61971142).

\bibliographystyle{model2-names}
\bibliography{AllBibliography_MSCMRSeg}

\end{document}